\documentstyle[preprint,aps]{revtex}
\begin{document}
\draft
\title{Exact Calculation of the Ground-State Dynamical
Spin Correlation Function
of a $S=1/2$ Antiferromagnetic Heisenberg Chain with Free Spinons}
\author{F. D. M. Haldane}
\address{Department of Physics, Princeton University, Princeton, New Jersey
08544}
\author{M. R.  Zirnbauer}
\address{Institut f\"{u}r Theoretische Physik,
Universit\"{a}t K\"{o}ln, Z\"{u}lpicherstr. 77, 50937 K\"{o}ln, Germany}
\date{August 17, 1993}
\maketitle
\begin{abstract}
We calculate the exact dynamical magnetic structure factor $S(Q,E)$
in the ground state of
a one-dimensional $S=1/2$ antiferromagnet with gapless free
$S=1/2$ spinon excitations, the Haldane-Shastry model with inverse-square
exchange, which is in the same low-energy universality class
as Bethe's nearest-neighbor exchange  model.
Only two-spinon excited states  contribute, and $S(Q,E)$ is found to
be a very simple
integral over these states.
\end{abstract}
\narrowtext
The integrable ``Haldane-Shastry'' model\cite{haldane88,shastry88}
(HSM) is a variant of the $S= 1/2$ Heisenberg chain, with exchange
inversely proportional to the square of the distance between spins,
in contrast to the nearest-neighbor exchange of the Bethe Ansatz
model\cite{bethe31} (BAM).  The low-energy properties of the antiferromagnetic
HSM are in the
same universality class as those of the BAM: the elementary excitations
are $S= 1/2$ objects\cite{faddeev} (``spinons'') obeying
{\it semion}
(half-fractional) statistics intermediate between bosons and
fermions\cite{haldane91fs},
and the low-energy fixed point is described
by the
$k=1$ $SU(2)$ Wess-Zumino-Witten model, a $c=1$ conformal field
theory\cite{haldane88} .
A  generic model in this universality class has marginally irrelevant
interactions that renormalize to zero at the infra-red fixed-point;
this is seen in the BAM.
The special feature of the HSM is
that such interactions
are absent
at {\it all } energy scales\cite{haldane91sg}:
it is in a very real sense the model
in which the spinons form  an ``ideal $S$ = 1/2 semion gas'', and may
be regarded as the fundamental model for gapless
half-integral spin antiferromagnetic chains.

While
the correlation functions of the BAM  have not yet been obtained,
the static (equal time) two-point antiferromagnetic
ground-state
spin-correlations of{the HSM are already known\cite{haldane88,shastry88}.
In this Letter, we extend this result to
the full {\it dynamical} spin correlation function:
\begin{equation}
\langle 0 | S^a_m(t) S^b_n(t') |0 \rangle = \case{1}{4}\delta^{ab}
(-1)^{m-n} C(m-n,t-t')
\end{equation}
where our expression for $C(x,t)$ is remarkably simple, and directly related
to the spinon spectrum. We will first present the result, then
describe the derivation.

The HSM Hamiltonian is
\begin{equation}
H = J \sum_{m<n} (d(m-n))^{-2}{\bf S}_m\cdot {\bf S}_n ,
\label{ham}
\end{equation}
where $d(n)$ = $(N/\pi)\sin (\pi n /N ) \rightarrow n $ as the number of sites
$N \rightarrow \infty$.  The HSM spinon dispersion relation is
\begin{equation}
 \epsilon (q) = (v /\pi ) \left ( (\pi /2)^2 - q^2 \right ),\quad v = \pi J/2 ,
\end{equation}
which is restricted to {\it half} the Brillouin zone
$(|q| < \pi/2 )$; $v$ is the low-energy spinon velocity.
For $N \rightarrow \infty$,
a state with $N_{sp}$ spinons has energy $\sum_i \epsilon (q_i) $ and
crystal momentum $\exp (iK) = (-1)^M \exp (i\sum_i q_i) $, where
$M$ = $(N-N_{sp})/2$ must be integral.

Our concise expression for $C(x,t)$ is
\begin{equation}
\int d{\bf M} \, \exp \left (\case{1}{4}i\pi  {\rm Tr}
[ x{\bf MM}_0- \case{1}{2} vt ({\bf 1} - ({\bf MM}_0)^2)]
\right ) .
\label{result}
\end{equation}
The integral
is over the manifold  of $4 \times 4 $ traceless Hermitian unitary
matrices ${\bf M}$ =
${\bf M}^{-1}$ =
$(\hat {\bf \Omega}_{L}\cdot \vec{\mbox{\boldmath$\sigma$}}_{L})
(\hat {\bf \Omega}_{R}\cdot \vec{\mbox{\boldmath$\sigma$}}_{R})$, where
$\{ {\mbox{\boldmath$\sigma$}}^a_L\}$ and  $\{ {\mbox{\boldmath$\sigma$}}^a_R
\}$
are two independent sets of $4 \times 4$ Hermitian
generators that individually obey the same
algebra as the Pauli matrices, and which commute with each other;
$\hat {\bf \Omega}_{L}$ and $\hat {\bf \Omega}_{R}$ are
real unit 3-vectors.
This manifold is isomorphic
to the product of two spherical surfaces $S^2_L \times S^2_R$.
The time-reversal operator for Pauli matrices is ${\mbox{\boldmath$\tau$}}$ =
$i {\mbox{\boldmath$\sigma$}}^2$,
so $({\mbox{\boldmath$\sigma$}}^a)^*$ =
${\mbox{\boldmath$\tau$}} {\mbox{\boldmath$\sigma$}}^a
{\mbox{\boldmath$\tau$}}$,
and ${\mbox{\boldmath$\tau$}}^2$ =
$-{\bf 1}$;
${\bf M}$ obeys the reality condition
${\bf M}^*$ = $({\mbox{\boldmath$\tau$}}_L {\mbox{\boldmath$\tau$}}_R) $
${\bf M}({\mbox{\boldmath$\tau$}}_L {\mbox{\boldmath$\tau$}}_R) $
where $({\mbox{\boldmath$\tau$}}_L{\mbox{\boldmath$\tau$}}_R)^2$ = ${\bf 1}$.
${\bf M}_0$ =
$(\hat{\bf z}\cdot \vec{\mbox{\boldmath$\sigma$}}_L)(\hat {\bf z} \cdot
\vec{\mbox{\boldmath$\sigma$}}_R)$
is a point on the manifold.
${\rm Tr}[{\bf M}{\bf M}_0 ]$ = $4\Omega_L^z\Omega_R^z$, and
${\rm Tr}[{\bf 1} - ({\bf M}{\bf M}_0)^2 ]$ = $8((\Omega_L^z)^2 +
(\Omega_R^z)^2 - 2(\Omega_L^z\Omega_R^z)^2 ) $.

The  invariant measure for the integral is
the product of rotationally-invariant
measures on  $S^2_L \times S^2_R$, and
the normalization is fixed so $C(0,0)$ = 1.
Writing  $\Omega_{L(R)}^z \equiv \lambda_{1(2)}$, this gives $C(x,t)$ as
\begin{equation}
 \case {1}{4} \int_{-1}^{1}d\lambda_1\int_{-1}^{1} d\lambda_2 \, e^{iQx-Et}
\end{equation}
where $Q = \pi \lambda_1\lambda_2$ and $E$ =
$(\pi v/2)(\lambda_1^2 + \lambda_2^2 -2\lambda_1^2 \lambda_2^2 )$.
Alternatively, if $\pi\lambda_1\lambda_2$ = $q_1 + q_2$, and
$(q_1-q_2)^2$ = $\pi^2 (1-\lambda_1^2)(1-\lambda_2^2)$,
so $Q = q_1 + q_2 $ and $E = \epsilon (q_1) + \epsilon (q_2 )$,
$C(x,t)$ is also given by
\begin{equation}
{v \over (2\pi )^2 }
\int_{-\pi/2}^{\pi/2} dq_1
\int_{-\pi/2}^{\pi/2} dq_2 \,
{ |q_1-q_2| \over  [\epsilon(q_1) \epsilon (q_2)]^{1/2}} e^{i(Qx-Et)} .
\label{phys}
\end{equation}
Note that the
$S=1$ state $S_n^{a} |0 \rangle $ is {\it completely} expressible
in terms of eigenstates of the HSM with {\it only two} parallel-spin spinons
carrying momenta $q_1$ and $q_2$.
Thus (\ref{phys}){\it explicitly
expresses the correlation function in terms of the
physical excitations}; this formula is the principal
result reported here.

$\langle 0 |S_m^a(t)S^b_n(t')|0\rangle $
can be expressed in terms of the {\it dynamical structure factor}
$S(Q,E)$ as
\begin{equation}
{\delta^{ab}\over 2\pi }\int_{0}^{2\pi}dQ\int_0^{\infty}dE \, S(Q,E)
e^{i(Q(m-n)-E(t-t'))} ,
\end{equation}
where $S(Q,E)$ is explicitly given by
\begin{equation}
\case{1}{4} \theta (E_2(Q) - E)
\prod_{\nu = \pm}
\left [{\theta (E- E_{1\nu}(Q))\over [E - E_{1\nu}(Q)]^{1/2} }\right ]
\label{sqe}
\end{equation}
where
$E_{1-}(Q) = (v/\pi)Q(\pi - Q) $,
$E_{1+}(Q) = (v/\pi)(Q - \pi)(2\pi - Q) $,
and $E_2(Q) = (v/2\pi)Q(2\pi- Q) $.
Integration over $E$ gives the previously-known expression for the
static structure factor:
$S(Q) =  \case{1}{4} \log (|1 - Q/\pi |^{-1})$ .

Our calculation  is based on the identification of $C(x,t)$ as
proportional to the {\it bosonic}
single-particle ground-state correlation function
$\langle 0 |\Psi^{\dagger}(x',t')\Psi(0,0)| 0 \rangle $
(with $x' \propto x $, $t' \propto t$)
for
the Calogero-Sutherland model\cite{CSM} (CSM)
of a spinless 1D Galilean-invariant gas with interactions
$(\hbar^2/4m)(\beta(\beta-2)/(x_i-x_j)^2)$,
at coupling constant $\beta$ = 4, the coupling at which
it is related to the symplectic random matrix ensemble\cite{dyson}.

Though different
in its technical details,
a recent calculation by Simons, Lee and Altshuler\cite{sla} (SLA) of
$\langle 0 | \rho (x',t')
\rho (0,0)|0\rangle $ for the $\beta = 4$ CSM
suggested our calculation.
Our result shows that the {\it hole} excitation in this model
fractionalizes into two semions (as in the essentially similar
bosonic Laughlin fractional quantum Hall effect (FQHE)
state at Landau level filling $\nu = 1/2$).
Our interpretation of the SLA expression\cite{sla} is that it  shows  that
the {\it particle} excitation remains elementary (the FQHE analog is
a particle added {\it outside} the Laughlin droplet);
inspection of  their formula  shows it
to be the convolution of our result for the fractionalized hole
contribution  with a
factor representing an  additional particle excitation
outside the ground state condensate.

We first note that (for finite $N$) the HSM exhibits not just
$SU(2)$ symmetry, with an associated $sl_2$ Lie algebra generated by
${\bf J}_0 = \sum_m {\bf S}_m$, but a remarkable
additional {\it quantum symmetry},
the {\it Yangian} $Y(sl_2)$ symmetry algebra with the additional
generator\cite{haldane92y}
\begin{equation}
 {\bf J}_1 = {h \over 2 }\sum_{m<n}
\cot \left ({\pi (m-n) \over N} \right) \,
{\bf S}_m \times {\bf S}_n
\label{Yangian}
\end{equation}
where $[H,J^a_m]$ =  $J^a_m |0 \rangle$ = 0, $m $ = 0,1.
Here $h$ is the ``quantum parameter'' of $Y(sl_2)$; this
is a scale parameter,
conventionally rescaled to 1.  However, in the limit
$N \rightarrow \infty$,
$hN $ must be held constant, so $h \rightarrow 0$.
The symmetry algebra then is
the ``classical'' infinite-dimensional Lie algebra
$\widehat{sl_2}_+$, which is the algebra of non-negative
modes of a Kac-Moody algebra:
$[J^a_m,J^b_n] = i \epsilon^{abc}J^c_{m+n}$, $m,n \ge 0$,
with $J^a_m |0 \rangle $ = 0, $m \ge 0$.

HSM eigenstates are organized into Yangian
``supermultiplets''\cite{haldane88,haldane92y}
(of many degenerate $SU(2)$ multiplets)
each containing a single Yangian
highest weight state  (YHWS) which is the state
of highest $J^z_0$ in the multiplet.  The set of YHWS
span a subspace of states that can be physically characterized as the
``fully-polarized spinon gas'' (FPSG) states\cite{haldane92y},
where all spinons have
maximally polarized spins.  The wavefunctions for FPSG
states can be conveniently expressed in terms of $M$ complex coordinates
$Z_i$ (with $(Z_i)^N = 1$) of lattice sites (on the unit circle)
with reversed spins:
\begin{equation}
\Psi(\{Z_i\}) = \Phi(\{Z_i\}) \prod_{i<j} (Z_i-Z_j)^2\prod_i Z_i ,
\end{equation}
where $\Phi(\{Z_i\})$ is a symmetric polynomial of degree
$N_{sp}$ in each variable; the constraint $ M = (N-N_{sp})/2 $
ensures that the Fourier transform of the wavefunction
has {\it compact support} within a {\it single} Brillouin zone.
Diagonalization of the HSM Hamiltonian within the
FPSG subspace gives the YHWS.
The other eigenstates are obtained from the YHWS by repeated
action of the Yangian generators, and the multiplets are classified
as irreducible representations of $Y(sl_2)$\cite{haldane92y,Bern}.
The structure of these multiplets is reminiscent of that of the
ideal $S=1/2$ Fermi gas, where there is an independent spin degree
of freedom for each orbital; $Y(sl_2)$ multiplets are also isomorphic
to direct products of independent $sl_2$ representations, but
in contrast to the fermion case, the ``orbital'' or ``Fock space''
structure {\it changes } with particle number.  These degeneracies
make it appropriate to identify the excitations of
the HSM as forming an ``ideal spinon gas''\cite{haldane91sg,haldane92y}.

The operator $S^+_m$ acting on a FPSG state
removes a down-spin coordinate, and preserves the FPSG character of
the state.  The wavefunction describing $S^+_m|0\rangle$
is
\begin{equation}
\prod_i (Z_i - z_m)^2 \prod_{i<j} (Z_i-Z_j)^2 \prod_i Z_i .
\end{equation}
The fact that the initial polynomial factor is degree-2 in
each coordinate $Z_i$ immediately shows that this state
is composed only of states with {\it two} spinon
excitations.

The polynomial character of the FPSG wavefunctions implies that
when matrix elements of the Hamiltonian between such states
are calculated, sums over
discrete lattice sites on the unit circle can be
replaced\cite{haldane88,haldane91sg} by
integrals of a continuous coordinate on the unit circle
(the compact support of the Fourier transforms of these
polynomials  within the first Brillouin zone
in reciprocal space
means that all ``Umklapp'' corrections vanish).  Thus calculations for the HSM
involving only FPSG states will give {\it identical} results to
analogous calculations for a continuous model, the $\lambda$ = 2
CSM on a circle.
Then
$C(x,t)$ $\propto$
$ \langle 0 | \Psi^{\dagger}(x',t') \Psi (0,0') |0\rangle $
where
$x'$ and $t'$ are space and time coordinates suitably scaled in terms
of CSM units.

Because of the special form of the HSM and CSM ground states,
$C(x,t)$ can also be written
\begin{equation}
\langle 0 | \prod_i (Z_i -z')^2
e^{-i(H-E_0)t'} \prod_i (Z_i - z)^2 | 0 \rangle .
\label{prod}
\end{equation}
We note that the method of our calculation can easily be extended to obtain
$\langle 0 | \prod_{j} \Psi^{\dagger}(x_j,t)
 \prod_{j} \Psi (x'_j,t') | 0 \rangle $
where  $n$  particles are removed at various places at one time
$t'$ and replaced at different places at
a different time $t$. (The final integration
over a manifold of $4n$-dimensional matrices will be more complicated,
however.)

In the thermodynamic limit,
the correlations of the CSM
for particles
confined in a weak harmonic potential that is rescaled
to maintain a constant particle density in some region
become locally equivalent to those of the same density
CSM model on a circle.
We use this Gaussian formulation,
since in this case\cite{matrix},  the dynamics of
the CSM particles become {\it identical} to the dynamics of the
doubly-degenerate eigenvalues of a
Hermitian matrix
with matrix elements that are harmonic oscillators
constrained so the matrix has Kramers degeneracy.
This is the {\it Gaussian Hermitian
matrix-model}\cite{matrix} with symplectic symmetry.

The action for the matrix-model is
\begin{equation}
S = \case{1}{2} \int dt \, {\rm Tr}
[ (\dot{{\bf Q}})^2 - \omega^2({\bf Q})^2 ]
\end{equation}
where ${\bf Q}$ is a $2N \times 2N $ Hermitian matrix
\begin{equation}
{\bf Q } = {\bf S} \otimes {\bf 1} + \sum_a {\bf A}^a \otimes
{\mbox{\boldmath$\sigma$}}^a .
\end{equation}
Here ${\bf S}$ is a real symmetric $N\times N$ matrix and
${\bf A}^a$ are three independent imaginary antisymmetric matrices,
making $2N^2 -N$ independent coordinates in total.
Then ${\bf Q}^* $ = ${\bf T}{\bf Q}{\bf T}^{-1} $, where
${\bf T}$ =
${\bf 1}\otimes {\mbox{\boldmath$\tau$}} $,
and ${\bf T}^2 = - {\bf 1 } $, which ensures Kramers degeneracy.
The limit
$N \rightarrow \infty $ with $\omega \propto 1/N $ is taken
at the end of the calculation.

The SLA calculation\cite{sla} was originally  a quenched
random matrix ensemble average, but can be reformulated\cite{slapre}
in terms of a
dynamical Gaussian Hermitian matrix-model, when it
is equivalent to a calculation of
$$\langle 0 |T_t \prod_{\alpha = \pm} {\rm Tr}
[( {\bf Q}(\alpha t) - \alpha x \pm i0^+ )^{-1}]
 | 0 \rangle $$
which is a
more complex calculation than the
one we describe here.

In (\ref{prod}), if the $\{Z_i \}$ are interpreted as the
doubly-degenerate eigenvalues of a unitary matrix ${\bf U}$ with Kramers
degeneracy belonging to Dyson's
circular ensemble with symplectic symmetry\cite{dyson},
$\prod_i (Z_i - z)^2$ $\equiv$ $\det [ {\bf U} - z ]$ .
Invoking the large-$N$ local equivalence of the eigenvalue
distributions of the circular unitary  and
Gaussian Hermitian ensembles\cite{dyson},
we identify the Gaussian
matrix-model quantity we need as
\begin{equation}
\langle 0 |T_t  \prod_{\alpha = \pm }
\det [{\bf Q}(\alpha t) -\alpha x] |0 \rangle
\end{equation}
where the space and time coordinates have been rescaled in terms of
matrix model units, and will be held fixed as $N \rightarrow \infty$.

The fundamental matrix-model correlation function is
\begin{equation}
\langle 0 |T_t {\rm Tr}[{\bf A}{\bf Q }(t) ]
{\rm Tr}[{\bf B}{\bf Q }(t') ]  | 0 \rangle
= G({\bf A},{\bf B})g(t-t')
\end{equation}
where
$g(t)$ = $(\hbar /2 \omega) \exp (i \omega |t| )$,
${\bf A}$ and ${\bf B}$ are $c$-number matrices,
and $G({\bf A},{\bf B})$ is
\begin{equation}
\case {1}{2}
\sum_{ij\sigma\sigma '}
\left (
A^{\sigma\sigma '}_{ij} B^{\sigma ' \sigma}_{ji}
+ A^{\sigma\sigma }_{ij} B^{\sigma ' \sigma '}_{ij}
- A^{\sigma\sigma '}_{ij} B^{\sigma ' \sigma}_{ij}
\right ) .
\end{equation}
The determinant is given by
a Gaussian Grassmann-number integral:
\[
\det [{\bf Q} (\alpha t)] = \int
\prod_{i\sigma } d\psi^*_{i\sigma}(\alpha )
d\psi_{i\sigma} (\alpha )
\exp {\rm Tr}[{\bf A } (\alpha ) {\bf Q } (\alpha t ) ]
\]
where
$A^{\sigma\sigma '}_{ij} (\alpha )$ =
$ \psi_{i\sigma}(\alpha ) \psi^*_{j\sigma '}(\alpha ) $.

We now evaluate the harmonic oscillator correlation
function, using the
standard harmonic oscillator result
$\langle 0 | T_t \exp A_1 \exp A_2 |0 \rangle $
= $\exp [(1/2) \sum_{ij} \langle 0 | T_t A_iA_j | 0 \rangle ] $
where $A_i$ are linear functions of the coordinates,
and obtain
\begin{equation}
\prod_{i\sigma\alpha }\int d\psi^*_{i\sigma}(\alpha)d\psi_{i\sigma}(\alpha)
\exp S[\{\Psi^*_{i\sigma}(\alpha),\Psi_{i\sigma}(\alpha ) \} ];
\end{equation}
here $S$ = $(xS_1 + (\hbar /8 \omega) S_2)$ where $S_1$ = $\sum_{\alpha}
\alpha \rho_{\alpha\alpha}$ and
$S_2$ = $\sum_{\alpha\beta} (g_{\alpha\beta}(t))^2
G({\bf A}(\alpha),{\bf A}(\beta) )$, with
\begin{equation}
G({\bf A}(\alpha),{\bf A}(\beta)) = - \left (
\rho_{\alpha\beta}\rho_{\beta\alpha} +
\case{1}{2} \phi^*_{\alpha\beta}\phi_{\beta\alpha}  \right ) ,
\end{equation}
and $g_{\alpha\beta}(t)$ = $\delta_{\alpha\beta} +
(1-\delta_{\alpha\beta})\exp(i \omega |t| )$. Here
$\rho_{\alpha \beta}$ = $ \sum_{i\sigma}
\psi^*_{i\sigma}(\alpha)\psi_{i\sigma}(\beta) $,
$\phi_{\alpha \beta }$ = $  \sum_{i\sigma\sigma '}
\tau_{\sigma\sigma'} \psi_{i\sigma}(\alpha)\psi_{i\sigma'}(\beta ) $
and
$\phi^*_{\alpha \beta }$ = $ \sum_{i\sigma\sigma '}
\tau_{\sigma\sigma'} \psi^*_{i\sigma}(\alpha)\psi^*_{i\sigma'}(\beta ). $

We now carry out a bosonic Hubbard-Stratonovitch transformation
to decouple $S_2$; ten independent real integration
variables are required, which
can be organized into a $2 \times 2$ Hermitian matrix $V_{\alpha\beta}$
coupling to $\rho_{\alpha\beta}$
and a $2 \times 2$ complex symmetric matrix $\Delta_{\alpha\beta}$
coupling to $\phi_{\alpha\beta}$, with $\Delta^*_{\alpha\beta}$
coupling to
$\phi^*_{\alpha\beta}$.
These can in turn be organized into a
$4 \times 4$  matrix ${\bf M}({\bf V},{\bf \Delta}, {\bf \Delta}^*)$ where
$M_{\alpha 1,\beta 1}$ = $V_{\alpha\beta}$,
$M_{\alpha 1,\beta 2}$ = $\Delta_{\alpha\beta}$,
$M_{\alpha 2,\beta 1}$ = $\Delta^*_{\alpha\beta}$,
and $M_{\alpha 2,\beta 2}$ = $V_{\beta\alpha}$.
Provided $V_{\beta\alpha}$ = $(V_{\alpha\beta})^*$
and $\Delta_{\alpha\beta}^*$ = $(\Delta_{\beta\alpha})^*$,
this matrix is Hermitian, and  obeys the reality condition
\begin{equation}
{\bf M}^* $ =   $({\bf 1}\otimes {\mbox{\boldmath$\sigma$}}^1 )
{\bf M} ({\bf 1}\otimes {\mbox{\boldmath$\sigma$}}^1 ) .
\label{reality}
\end{equation}

Such a matrix has the ten-parameter form
\begin{equation}
{\bf M} = C^0 {\bf 1} + \sum_{ab}
C^{ab}{\mbox{\boldmath$\sigma$}}^a_L{\mbox{\boldmath$\sigma$}}^b_R .
\end{equation}
where $C^0 $ is a real number and ${\bf C}$ is a real $3 \times 3$
matrix.
The explicit realization of the generators is
$\vec{\mbox{\boldmath$\sigma$}}_L$ = $(
{\mbox{\boldmath$\sigma$}}^2 \otimes {\mbox{\boldmath$\sigma$}}^2,
-{\mbox{\boldmath$\sigma$}}^2 \otimes {\mbox{\boldmath$\sigma$}}^1,
{\bf 1} \otimes {\mbox{\boldmath$\sigma$}}^3)$
and  $\vec{\mbox{\boldmath$\sigma$}}_R$ = $ (
{\mbox{\boldmath$\sigma$}}^1 \otimes {\mbox{\boldmath$\sigma$}}^3,
{\mbox{\boldmath$\sigma$}}^2 \otimes {\bf 1},
{\mbox{\boldmath$\sigma$}}^3 \otimes {\mbox{\boldmath$\sigma$}}^3)$.
The unitary transformations generated by
$\{{\mbox{\boldmath$\sigma$}}^a_L,{\mbox{\boldmath$\sigma$}}^a_R \}$ correspond
to a
$G = SO(3)_L \otimes SO(3)_R $ continuous symmetry group,
under the action of which
$ {\bf C} \rightarrow {\bf O}_L {\bf C} {\bf O}_R^{-1} $.

We now  take the large-$N$ limit with $\omega \propto 1/N$,
and $x',t' \sim O(1)$.
The Grassmann integrals decouple into a product of $N$ identical
integrals leading to a factor
\begin{equation}
\left ( \det [{\bf M} + iN^{-1}\delta {\bf M} ] \right )^N
\end{equation}
where (dropping scale factors in $x'$ and $t'$)
\begin{equation}
\delta {\bf M} = (x' - t'[{\bf M},{\bf M}_0]){\bf M}_0 ,
\end{equation}
${\bf M}_0 $ = $({\mbox{\boldmath$\sigma$}}^3 \otimes {\bf 1} )$.
The correction $\delta {\bf M}$ shifts the matrix in the
determinant resulting from the Grassmann integrals away from
the Hermitian value ${\bf M}$ of the Hubbard-Stratonovich fields, but
this does not affect the validity or convergence of the Grassmann
integral.  Using the identity $\det [{\bf M}]$
$\equiv$ $ \exp ({\rm Tr}[{\rm ln}{\bf M}])$, the value of the integrand
for large $N$, with the Gaussian Hubbard-Stratonovich factor included,
becomes
$( \det [{\bf M} ]\exp ( -{\rm Tr}[{\bf M}^2]/2)) ^N \exp (i\varphi ) $
where $\varphi$ =
${\rm Tr }[ {\bf M}^{-1} (x' - t'[{\bf M},{\bf M}_0]){\bf M}_0 ]$ .
The factor which is exponentiated to a power $N$ is maximized when
${\bf M}^2 $ = ${\bf 1}$, or ${\bf M}^{-1} $ = ${\bf M}$,
so at the  extremal points,
\begin{equation}
\varphi = {\rm Tr}[ x'{\bf M}{\bf M}_0 - t'({\bf 1}- ({\bf M}{\bf M}_0)^2)] .
\end{equation}

There are five classes of solutions of the condition
${\bf M}^2$ = 1 corresponding  to the five possible
signatures $s$ = $\pm 4$, $\pm 2$, and 0 of ${\bf M}$, all with
${\rm Tr}[{\bf M}]$ = $s$ and
$C^0 = s/4$.  For $s= \pm 4$, there are two discrete
solutions with ${\bf C} $ = 0, invariant under the
full group $G$.    For $s$ = $\pm 2$, ${\bf C}$ = $C^0 {\bf O}$, where
${\bf O}$ is orthogonal, which is invariant under the
$SO(3)$ subgroup of $G$ where ${\bf O}_L$ = ${\bf O} {\bf O}_R
{\bf O}^{-1}$, giving two manifolds of dimension 3.  Finally, for
$s$ = 0, $C^{ab}$ = $\Omega^a_L\Omega^b_R$ where
$\hat{\bf \Omega}_{L(R)}$ are unit 3-vectors; this is invariant
under the $O(2)_L\otimes O(2)_R$ subgroup of $G$
where $O_{L(R)}$ is a rotation around $\hat{\bf \Omega}_{L(R)}$,
and forms a manifold of dimension 4.

When the same extremal value occurs on several different manifolds,
only manifolds with the largest dimension contribute to the
integral as $N \rightarrow \infty$.
The measure for the integral over the dominant $s=0$ extremal manifold is
just the invariant measure
on $S^2_L \times S^2_R$.
To complete the derivation of (\ref{result}),
the space and time variables $x'$ and $t'$
must be rescaled into HSM units.
This is accomplished by noting that
the two-spinon states span a momentum range of $2 \pi $ and
an energy range of $\pi v/ 2$.

Finally, we comment on the relation between our HSM result
(\ref{phys}) and the BAM correlations.
An {\it Ansatz} similar to the exact HSM result (\ref{sqe}) for $S(Q,E)$
has previously been proposed\cite{mueller}
as an {\it approximation} to the BAM structure factor.
The key simplification in the HSM is that only the two-spinon
excited states contribute to $S(Q,E)$ at $T = 0$.  In the BAM,
the numerical studies\cite{mueller} show that while these states
dominate the spectral weight, there is a finite contribution from
the $N_{sp}  > 2 $  states which, while  small,
presumably remains finite for any large but finite $N_{sp}$.

A variant of the HSM with $N = \infty$ and $d(n)$
= $ \kappa^{-1}\sinh (\kappa n ) $, $\kappa$ real,
has been identified\cite{into}
as a family of integrable models that interpolate between the
$N=\infty$ HSM ($\kappa = 0$)  and the BAM ($\kappa = \infty$).
These models also have a $Y(sl_2)$ quantum
symmetry\cite{haldane92y} generated by the analog of (\ref{Yangian})
with $\cot (\pi (m-n)/N) $ replaced by $i \coth (\kappa (m-n) $;
this realization of $Y(sl_2)$ goes over smoothly into its more-familiar
realization in the $N=\infty$ BAM limit.

The Yangian quantum parameter $h$ must  be rescaled to zero to keep
$h /\kappa $ constant in the
HSM limit $\kappa \rightarrow 0$ with $N = \infty$.
The reverse process by which the ``classical''
$\widehat{sl_2}_+ $ symmetry at $\kappa = 0$
becomes the
$Y(sl_2)$ ``quantum'' symmetry when $\kappa > 0 $ is
a {\it ``quantum deformation''}. We speculate that the extension of
the  $\kappa = 0 $ result (\ref{phys}) to the $\kappa > 0 $
model\cite{into}, and
thence to the $\kappa = \infty $ BAM limit, may be achievable as
such a ``quantum deformation''.
In this scenario, there is
an expression for
$C(m,t;\kappa)$  as an expansion in multi-spinon terms
with $N_{sp} = 2,4,6,\ldots ,\infty$;
when $\kappa = i\pi /N$ the $N_{sp} > 2$
terms would vanish, giving the  finite-$N$ HSM correlations.

This work was conceived and carried out at the Aspen Center
for Physics during its 1993 Summer Program.
MRZ acknowledges partial support
from the Sonderforschungsbereich 341, K\"{o}ln-Aachen-J\"{u}lich.

\end{document}